\documentstyle[12pt]{article}
\def\beq{\begin{equation}}
\def\eeq{\end{equation}}
\def\bea{\begin{eqnarray}}
\def\eea{\end{eqnarray}}
\def\bq{\begin{quote}}
\def\eq{\end{quote}}
\def\PRL{{\it Phys. Rev. Lett.} }
\def\PL{{\it Phys. Lett.} }
\def\NP{{\it Nucl. Phys.} }
\def\PR{{\it Phys. Rev.} }
\def\AJ{{\it Astrophys. J.}}
\def\CMP{{\it Commun. Math. Phys.} }
\def\AM{{\it Ann. Math.} }
\def\BAMS{{\it Bull. Am. Math. Soc.} }
\def\MPL{{\it Mod. Phys. Lett.} }
\def\HPA{{\it Helv. Phys. Acta} }
\def\AP{{\it Ann. Phys.} }
\def\JP{{\it J. Phys.} }
\def\IJMP{{\it Int. J. Mod. Phys.} }
\def\RMPA{{\it Rev. Math. Pure Appl.} }
\def\CMJ{{\it Czech. Math.J.} }
\def\JMM{{\it J. Math. Mech.} }

\def\gappeq{\mathrel{\rlap {\raise.5ex\hbox{$>$}}
{\lower.5ex\hbox{$\sim$}}}}

\def\lappeq{\mathrel{\rlap{\raise.5ex\hbox{$<$}}
{\lower.5ex\hbox{$\sim$}}}}
\begin{document}
\renewcommand{\thesection}{\Roman{section}.}
\renewcommand{\theequation}{\arabic{section}.\arabic{equation}}
\topmargin -0.5cm
\oddsidemargin -0.8cm
\evensidemargin -0.8cm
\pagestyle{empty}
\begin{flushright}
CERN-TH.7167/94
\end{flushright}
\vspace*{3mm}
\begin{center}
{\bf UNIFICATION OF FUNDAMENTAL INTERACTIONS IN SUPERSYMMETRY
\footnote{ Invited Talk at the Conference on ``Unified Symmetry in the
Small and in the Large", at Coral Gables,\newline \phantom{~~~1}
Florida, Jan
27--30, 1994.}}\\ \vspace*{1cm} Pran Nath\\
\vspace{0.3cm}
Theoretical Physics Division, CERN\\
CH-1211 Geneva 23, Switzerland\\
and\\
Department of Physics, Northeastern University\\
Boston, MA 02115, USA\footnote{Permanent address.}\\
\vspace*{0.5cm}
and\\
\vspace*{0.5cm}
R. Arnowitt\\
\vspace{0.3cm}
Center for Theoretical Physics, Department of Physics\\
Texas A\&M University, College Station, TX 77843, USA\\
\vspace*{1cm}
Abstract
\end{center}
A review is given of recent developments on the implications of
supergravity grand
unification with SU(5)-type proton decay under the condition
$M_{H_3}/m_G < 10$
(where $M_{H_3}$ is the Higgs triplet mass) and the naturalness
condition that the
universal scalar mass $m_0$ and the gluino mass are $<$ 1~TeV.  It is
shown that the
maximum achievable lifetime limits on proton lifetime at Super
Kamiokande and
ICARUS will exhaust the full parameter space of the model under the
constraint
$m_{\tilde W_1} >$~100~GeV.  Thus the model predicts the observation of
either a
light chargino with mass $\lappeq$~100~GeV, or the observation of a
$\bar \nu K^+$
mode at Super Kamiokande and ICARUS within the above naturalness
constraints.
Analysis of the $b \rightarrow s\gamma$ branching ratio within this
model is also
discussed.  It is shown that there is a significant region of the
parameter space
where the branching ratio predicted by the model lies within the current
experimental bounds.  It is pointed out that improved measurements of
$B(b
\rightarrow s\gamma)$ will significantly delineate the parameter space
of the
model and allow for a more stringent determination of their allowed
ranges.

\vspace*{0.7cm}

\begin{flushleft}
CERN-TH.7167/94\\
February 1994
\end{flushleft}
\vfill\eject
\pagestyle{empty}
\clearpage\mbox{}\clearpage
\setcounter{page}{1}
\pagestyle{plain}

\section{\sc Introduction}

One of the important developments over the past three years has been the
demonstration that the high precision LEP data \cite{aaa}, when
extrapolated to
high energy using renormalization group equations, leads to unification
of the
SU(3)$_{\rm C}~\times$ SU(2)$_{\rm L}~\times$ U(1) coupling constants
$\alpha_3,
\alpha_2$ and $\alpha_1$ (where $\alpha_1 = \frac{5}{3} \alpha_Y$ and
$\alpha_Y$
is the hypercharge coupling constant) within the standard supersymmetric
SU(5)
theory \cite{bb}.  Supergravity grand unification provides an attractive
mechanism for the unification of the electroweak and the strong
interactions as
well as a framework where supersymmetry can be broken in a consistent
fashion
\cite{cc},\cite{dd}.  Thus the spontaneous breaking of supersymmetry in
unified
supergravity models leads to an enormous reduction
\cite{cc}, \cite{ee}--\cite{ggg} in the number of arbitrary
SUSY-breaking
parameters that one encounters in globally supersymmetric theories
\cite{hh}.

We begin by reviewing briefly the ideas of supergravity grand
unification.  Our
starting assumption is the existence of an $N=1$ supergravity unified
theory
below the Planck scale $M_{Pl} = 2.4 \times 10^{18}$~GeV.  This unified
theory
can be completely specified in terms of three functions.  These are: (1)
superpotential $W(z_a)$, which is a holomorphic function of the chiral
fields, (2)
a K\"ahler potential $K(z_a,z^*_a)$, which depends both on the chiral
fields and
on their complex conjugates, and (3) a gauge kinetic energy function
$f_{\alpha\beta}$, which also is a function of the fields and their
complex
conjugates.  One obtains a gaugino mass from supergravity couplings,
which is of
the form
\beq
\frac{1}{4} \bar e^{\frac{G}{2}}(G^{-1})^a_b~G,^{~a}~f^*_{\alpha\beta,b}
\bar\lambda^\alpha\lambda^\beta
\label{1}
\eeq
where $G = -\ln [\kappa^6WW^*] - \kappa^2K$, and $\kappa=
(M_{Pl})^{-1}$.  We
shall make the standard assumptions that appear in the formulation of
superunified models.  First we shall assume that supersymmetry is broken
by a
gauge singlet field in the hidden sector \cite{cc},\cite{ee}.  Next, we
shall
assume that the K\"ahler potential is generation-blind at the GUT scale
so that
FCNC will be suppressed at low energies, and that the superpotential is
restricted to renormalizable interactions (although an understanding of
the full
quark and lepton mass hierarchy may require inclusion of higher
dimensioned
operators).  With the above assumptions, one can generate a low energy
effective
theory below the GUT scale, where the gauge group is SU(3)$_{\rm C}
\times$
SU(2)$_{\rm L} \times$ U(1)$_{\rm Y}$ and the superheavy sector is
integrated
out.  The effective potential of this theory is given by $V_{eff} =
V_{SS} +
V_{SB}$ where $V_{SS}$ is the supersymmetry-invariant part and $V_{SB}$
is the
symmetry-breaking part.  In supergravity a very simple form emerges for
$V_{SB}$:
\beq
V_{SB} = \sum_i m^2_0z_iz^+_i + \left( A_0W^{(3)} + B_0W^{(2)} +
M^{-1}_{H_3}W^{(4)} \right)~,
\label{2}
\eeq
where $W^{(2)}$ is the quadratic, $W^{(3)}$ is the cubic and $W^{(4)}$
the
quartic part of the effective superpotential below the GUT scale.  After
SUSY
breaking, Eq.~(\ref{1}) gives a gaugino mass term of the form
$m_{\frac{1}{2}}\bar\lambda^\alpha \lambda^\alpha$, so that together
with
Eq.~({\ref{2}) one has a total of four soft SUSY-breaking parameters
\beq
m_0,~m_{\frac{1}{2}},~A_0,~B_0~.
\label{3}
\eeq

A phenomenologically  interesting GUT group in general could be an
SU(5), SO(10),
E(6), etc.  We shall assume that the specific model we are considering
is or
contains an SU(5).  For the spectrum, we shall assume three generations
of quarks
and leptons in 5+$\overline {10}$ representations, and two plets of
Higgs
$(H_1,H_2)$ in (5,$\overline{5}$) representations.  At low energy, the
effective
superpotential will then have a quadratic term of the form $W^{(2)} =
\mu_0H_1H_2$,
where $H_1, H_2$ now refer to the Higgs doublets in the respective
multiplets.

The outline of the rest of the paper is as follows:  in Section 2, we
discuss the
analysis using radiative electroweak symmetry breaking under a number of
constraints which are physically desirable.  In Section~3, we discuss
the
question whether Super Kamiokande and ICARUS can test SUGRA GUTs.  In
Section~4,
we discuss $b \rightarrow s \gamma$ decay in SU(5) SUGRA GUTs.
Conclusions are
given in Section~5.

\section{\sc Analysis using radiative electroweak symmetry\newline
breaking}
\setcounter{equation}{0}

As is well known an attractive feature of supergravity grand unification
is that
it can generate its own electroweak symmetry breaking via
renormalization group
effects \cite{jj}.  Under the assumption of charge and colour
conservation, the
potential that governs symmetry breaking in supergravity models is given
by $V =
V_0 + \Delta V_1$, where $V_0$ is the renormalization group improved
semi-classical
tree potential
\bea
V_0(Q) &=& m^2_1(t) |H_1|^2 + m^2_2(t)|H_2|^2 - m^2_3(t)(H_1H_2 + {\rm
h.c.})
\nonumber \\
&+& \frac{1}{2} (g^2_2 + g^2_Y)(|H_1|^2 - |H_2|^2)^2
\label{21}
\eea
and $\Delta V_1$ gives the one-loop correction \cite{kk}$-$\cite{mm}
\beq
\Delta V_1 = \frac{1}{64\pi^2} \sum_i(-1)^{2j_i+1} n_i \left[ M^4_i
\left(\log
\frac{M^2_i}{Q^2} - \frac{3}{2} \right) \right]~.
\label{22}
\eeq
The analysis of electroweak symmetry breaking is carried out by what has
now
become a standard procedure.  One evolves the renormalization group
equations on
gauge and Yukawa couplings and on soft SUSY-breaking terms.  In addition
to the
charge and colour conservation, one imposes CDF and LEP lower bounds and
naturalness upper limit of 1~TeV on $m_0$ and $m_{gluino}$.

Additionally, in supergravity GUTs one imposes the proton lifetime lower
limits
from Kamiokande/IMB experiments.  The radiative breaking analysis
\cite{nn}
determines the parameter $\mu_0$ by fixing $M_Z$, and the parameter
$B_0$ can be
traded in favour of $\tan \beta$.  Thus the theory may be described by
the four
parameters
\beq
m_0~,~~ m_{\frac{1}{2}}~,~~A_0~,~~\tan \beta~.
\label{23}
\eeq
There are 32 supersymmetric particles in the theory, whose masses can be
computed
in terms of the four parameters of Eq.~(\ref{23}).  Thus there are 28
predictions
relating the SUSY masses in supergravity unification
\cite{nn}$-$\cite{zz}.

Some of the main results of the analysis described above are now
discussed:
\begin{itemize}
\item[(1)] Scaling laws \cite{nn}$-$\cite{pp}:  For a large part of the
parameter
space one finds $\mu \gg M_Z$, and scaling laws emerge.  Specifically
for the
neutralino and chargino masses one finds the relation
$$
2m_{\tilde Z_1} \simeq m_{\tilde W_1} \simeq m_{\tilde Z_2}
\eqno{(2.4a)}
$$
$$
m_{\tilde Z_3} \simeq m_{\tilde Z_4} \simeq m_{\tilde W_2} \simeq |\mu|
\eqno{(2.4b)}
$$
$$
m_{\tilde W_1} \simeq \frac{1}{4} m_{\tilde g} (\mu > 0),~~m_{\tilde
W_1} \simeq
\frac{1}{3} m_{\tilde g} (\mu < 0)~.
\eqno{(2.4c)}
$$
 Similarly the four Higgs bosons, except
the lightest Higgs $(h^0)$, are seen to have essentially degenerate
masses:
 \addtocounter{equation}{1}
\beq
m_{H^+} \simeq m_{H^0} \simeq m_A~.
\label{25}
\eeq
\item[(2)] Limits on the light Higgs and the top \cite{nn},\cite{oo}:
The analysis
gives upper limits that are
$$
m_{h^0} \lappeq 110{\rm -}120~{\rm GeV}
\eqno{(2.6a)}
$$
$$
m_t \lappeq 180{\rm -}190~{\rm GeV}~.
\eqno{(2.6b)}
$$

\item[(3)] Other spectra:  Radiative electroweak symmetry breaking and
proton
stability coonstrain other SUSY mass spectra as well.  The proton
stability
constraint can be easily understood in a qualitative fashion in the
limit when
$m_0$ is large.  This limit gives, for the dressing loop function $B$
[see
Eq.~(\ref{32})], the result
 \addtocounter{equation}{1}
\beq
B \simeq -2 \left( \frac{\alpha_2}{\alpha_3} \sin 2 \beta \right)
\left( m_{\tilde g}/ m^2_{\tilde q} \right)
\label{27}
\eeq
and the approximate relation $m^2_{\tilde q} = m^2_0 + 0.65 m^2_{\tilde
q}$.  Thus
a small gluino mass, a large squark mass and a small $\tan \beta$ are
favoured by
proton stability.  Typically this implies that the first two generations
fo squarks
will be essentially degenerate in mass and heavier than the gluino.
Similarly,
masses of the three generations of sleptons will be essentially
degenerate and
large, but lighter than the first two generations of squarks.  Masses of
the third
generation of squarks is more complicated, because a heavy top mass can
lead to a
very small mass for the lighter of the two stop masses.  In fact the
condition that
the stop masses not turn tachyonic acts as a constraint on the parameter
space
of the model.  Proton stability constraints also give a lower limit on
the mass of
the $A$-Higgs boson, which is significantly larger than what is obtained
from
electroweak symmetry breaking alone.  In this context we recall that
there exists a
hole in the parameter space of the MSSM which extends roughly from 100
to 200~GeV
(with $\tan \beta$ in the range $5 \lappeq \tan \beta \lappeq 20)$ which
cannot be
explored by LEP2 and LHC experiments \cite{aai}.  The constraints of
radiative
electroweak symmetry breaking with proton stability already give a lower
bound on
the $A$-Higgs mass that closes this gap.
 \end{itemize}

\section{\sc Can the sugra gut be tested at super kamiokande and
icarus?}
\setcounter{equation}{0}
There already exist stringent limits on the proton lifetime, and these
limits
are expected to improve significantly in the new generation of proton
stability experiments, i.e. Super Kamiokande and ICARUS.  One may ask if
the
expected increase in the sensitivity of these proton decay experiments
will be
able to test in a conclusive fashion at least a class of SUGRA GUT
models.  To
quantify the discussion, we shall assume that the GUT group $G$ we are
dealing
with is or contains an SU(5).  We assume further that at the GUT scale,
the GUT
group $G$ breaks into the Standard Model gauge group SU(3)$_{\rm
C}~\times$
SU(2)$_{\rm L}~\times$ U(1)$_{\rm Y}$.  We also assume the existence of
just
two doublets of Higgs, which are embedded in 5+$\overline{5}$ of SU(5).
Finally, we assume that there are no discrete symmetries in the model
which
forbid proton decay.  We shall focus here on the dominant decay mode of
$p
\rightarrow \bar \nu K^+$.  One can easily show that there is a
model-independent amplitude for the above mode which goes via the
exchange of
the colour triplet Higgsino field $\tilde H_3$, and one may write the
decay width as
\beq
\Gamma (p \rightarrow \bar \nu K^+) = {\rm Const} \left(
\frac{\beta_p}{m_{H_3}} \right)^2 |B|^2~.
\label{31}
\eeq
In Eq.~(\ref{31}), the Const is a product of phase-space and chiral
Lagrangian
factors \cite{bbi}, $\beta_p$ is the three-quark matrix element of the
proton,
whose value is known from lattice gauge theory, $M_{H_3}$ is the
Higgsino
triplet mass and $B$ is the dressing loop function defined in
Refs.~\cite{nn}--\cite{oo}.  Using the current experimental lower limit
on this
mode, which is \cite{cci} $\tau(p \rightarrow \bar\nu K^+) > 1.0 \times
10^{32}~{\rm yr}$, one finds on using Eq.~(\ref{31}) an upper limit of
\beq
B \lappeq 100(M_{H_3}/M_G)~{\rm GeV}^{-1}~,
\label{32}
\eeq
where we expect $M_{H_3}/M_G \lappeq 10$.  In the future, Super
Kamiokande and
ICARUS are expected to reach lower limits of
$$
{\rm Super~Kamiokande}~\cite{ddi} ~ \tau (p \rightarrow \bar\nu K^+) > 2
\times
10^{33}{\rm yr}
\eqno{(3.3a)}
$$
$$
{\rm ICARUS}~\cite{eei}~ \tau(p \rightarrow \bar\nu K^+) > 5 \times
10^{33}{\rm yr}
\eqno{(3.3b)}
$$
In view of the increased sensitivity expected in these experiments, one
may ask
if Eq.~(3.3) will exhaust the full parameter space of the SU(5)-type
SUGRA
GUT.  A detailed analysis of this question shows \cite{ffi} that
Eq.~(3.3) would not itself be able to exhaust the full parameter space
of
the SU(5)-type GUT.  However, it was found that Super Kamiokande and
ICARUS
experiments, along with maximum achievable superparticle mass limits at
LEP2
and at the Tevatron can exhaust the full parameter space \cite{ffi}.
Specifically one finds the following results \cite{ffi}:
\begin{itemize}
\item[(a)]
If $\tau (p \rightarrow \bar\nu K^+) > 1.5 \times 10^{33}{\rm yr}$, then
either
$m_{h^0} \lappeq 95$~GeV or $m_{\tilde W_1} < 100$~GeV.  Thus either
$h^0$ or
$\tilde W_1$ (and possibly both) will be observable at LEP2, provided
LEP2 can reach
its optimum energy and luminosity.

\item[(b)] Either the $\bar\nu K^+$ mode should be seen at the
Super Kamiokande and ICARUS experiments, or the $\tilde W_1$ should be
seen at
LEP2.
\end{itemize}

The result of case (b) above is exhibited in Fig.~1, where the maximum
value of
$\tau(p \rightarrow \bar\nu K)$ is given when $m_{\tilde W_1} >
100$~GeV, for the
case $\mu > 0$ and $m_t = 150$~GeV.  The maximum lifetime at a given
$m_0$ is
obtained by exhausting the allowed domain in the rest of the parameter
space,
i.e. $m_{\tilde g}, A_t, \tan \beta$.  Figure~1 shows that with
$M_{H_3}/M_G <
10$, ICARUS will exhaust the entire parameter space of the SU(5)-type
SUGRA GUT
under the restriction that $m_{\tilde W_1} > 100$~GeV.  Thus the
conclusion of (b)
above follows.  The above analysis shows that the SU(5)-type SUGRA GUT
can be tested
by using a combination of accelerator and non-accelerator experiments.

\section{\sc $b \rightarrow s\gamma$ decay in sugra gut}
\setcounter{equation}{0}
Recently CLEO \cite{ggi} has obtained a new upper bound on the inclusive
process
$b \rightarrow s \gamma$ with a value $BR(b \rightarrow s\gamma) < 5.4
\times
10^{-4}$ [at 95\% CL].  They also observe a non-vanishing result for the
exclusive
mode $B \rightarrow K^*\gamma$ with a branching ratio of $5 \times
10^{-5}$.
Assuming that the $K^*\gamma$ contributes $\approx 15\%$ to the
inclusive
process, as is indicated by lattice gauge calculations \cite{hhi}, one
obtains also a
lower limit $BR(b \rightarrow s \gamma) > 1.5 \times 10^{-5}$.  The
branching
ratio measurements on $ b \rightarrow s\gamma$ are expected to improve
in the
future, both with analysis of additional data at CLEO and from data that
would emerge from $B$-factories, where one expects luminosities that
would be an
order of magnitude larger than at current machines.  The Standard Model
result
would then be put to a severe test, and any deviation from it would
signal the onset of new physics beyond the SM.

In the ${\rm SM}, b \rightarrow s\gamma$ proceeds  via a penguin, which
involves
exchange of a $W$ boson and gives $BR (b \rightarrow s \gamma) \simeq
3.5 \times
10^{-4}$.  This result is ambiguous up to few per cent since the $BR$ is
a sensitive
function of the inputs such as quark masses and $\alpha_3$, which
currently have some
inherent experimental errors.  Additionally, the current analyses of the
$BR$ are
done only to leading order QCD corrections, and there may be important
beyond the
leading-order corrections which have not yet been fully computed
\cite{jji}.
There are additional penguins in supergravity which contribute to this
branching
ratio.  These involve the exchange of charged Higgses, charginos,
gluinos and
neutralinos \cite{kki}--\cite{mmi}.  We summarize here results of a
recent
analysis of these contributions within the framework of electroweak
symmetry
\cite{nni},\cite{ooi} and SUGRA GUT constraints including the constraint
of proton
stability \cite{nni}.

In the SUGRA GUT analysis here we compute the contributions from $W$,
charged
Higgses and charginos and neglect small contributions from the
neutralino and gluino
exchanges.  To leading order QCD corrections, one has
\cite{ppi},\cite{kki},\cite{lli}
\beq
\frac{BR (b \rightarrow s \gamma)}{BR (b \rightarrow ce\nu)} =
\frac{6\alpha}{\pi}\,\, \frac{\left[\eta^{\frac{16}{32}}A_\gamma +
\frac{8}{3}
\left( \eta^{\frac{14}{23}} - \eta^{\frac{16}{32}}A_g \right) + C
\right]^2}
{P \left(\frac{m_c}{m_b} \right) \left[1 - \frac{2}{3\pi} \alpha_s(m_b)
\left( \frac{m_c}{m_b} \right) \right]}
\label{41}
\eeq
where $\eta = \alpha_s(m_Z) / \alpha_s(m_b), f(m_c/m_b) = 0.241$, $C$ is
an
operator mixing coefficient and $P$ is a phase space factor.  For $C$
we use the recent analysis of Ref.~\cite{qqi}.

For the case of the SUGRA SU(5) GUT one finds \cite{nni} that there is a
significant region of the parameter space where the branching ratio $BR
(b
\rightarrow s \gamma)$ lies within the current experimental bound.

Results for the case $\mu >  0$ and $m_t = 150$~GeV are exhibited in
Fig.~2.  The
vertical line at $B = 1000$~GeV$^{-1}$ gives the current value of the
Kamiokande
experimental bound with $m_{H_3}/M_G$ = 10.  The region to the right of
this
vertical line is thus forbidden by $p$-stability while the region to the
left is
allowed.  We see that the branching ratio can be either larger and
smaller than the
SM value in the allowed domain, with significant deviations from the SM
value
occurring in both directions.  The analysis of the $\mu < 0$ case,
although
quantitatively different, is similar to the $\mu > 0$ case.  Again one
finds here a
significant region of the parameter space where the model gives a $b
\rightarrow
s\gamma$ branching ratio within the current experimental bounds.

The $b \rightarrow s \gamma$ branching ratio is also a sensitive
function of the
hidden sector parameter $A_t$, and of $\tan\beta$. In SUGRA SU(5) GUT
$\tan
\beta$ is limited by $p$-stability in such a way that $\tan\beta < 7-8$.
If the
constraint of $p$-stability is eliminated, $\tan \beta$ becomes
unrestricted and
much larger variations in the $b \rightarrow s \gamma$ branching ratio
can be
generated \cite{ooi}.

$b$--$\tau$ unification:  Recent analyses on $b$--$\tau$ masses within
SUSY
SU(5)-type models point to rather stringent constraints on $\tan \beta$.
Specifically it is found that the constraint of the equality of the
$b$--$\tau$
Yukawa couplings at the GUT scale implies that the top mass lies close
to its fixed
point value \cite{rri}.  The analyses reveal two branches in $\tan
\beta$ for fixed
$m_t$.  For the range of the top mass in the LEP favoured region of
130--170~GeV,
$\tan \beta$ is found to be either small $(< 2)$ or very large $(> 50)$.
These
constraints on $\tan\beta$ are rather severe, and it is reasonable to
ask how rigid
they are.  Before a fixed conclusion can be drawn, it is necessary to
carry out a
re-evaluation of the inputs used in the analyses.  These include the $b$
and $\tau$
masses, $\alpha_3$, as well as specific assumptions made in the
renormalization
group analyses, such as the treatment of the low energy and GUT
thresholds.  Aside
from the above there is the philosophical issue of whether it is
reasonable to
impose a strict $b$--$\tau$ unification since $s$--$\mu$ and $d$--$e$
unification are
not nearly as good.  This latter feature points to the possibility of
higher-dimensional operators or Planck slop terms at the GUT scale.  The
existence
of such operators appears reasonable in view of the fact that we are
dealing with
an effective theory, even at the GUT scale, and there may be quantum
gravity
corrections induced at this scale due to new physics at the Planck
scale.
Inclusion of such Planck slop terms via higher-dimensional
operators indicates a loosening of the stringent constraints discussed
above
\cite{ssi}.

\section{\sc conclusions}

Supergravity SU(5) GUT is a leading contender for the unification of
electroweak
and strong interactions.  The model is presently consistent with all
known
experiment, and makes predictions that are accessible at current
accelerator and
non-accelerator experiments, and at machines that are expected to go on
line in
the near future.  Specifically in this review, it was shown that
although
Super Kamiokande and ICARUS by themselves cannot exhaust the full
parameter space
of SU(5)-type supergravity GUTs, they can do so when combined with the
maximum
achievable superparticle mass limits at LEP2 and the Tevatron.  An
analysis of the
$b \rightarrow s\gamma$ decay in the SU(5) supergravity GUT was also
given.  It is
found that there is a significant region of the parameter space where
the $BR(b
\rightarrow s\gamma)$ predicted by the theory lies within the current
experimental
bounds.  It is pointed out that improved experimental limits may be able
to put more stringent constraints on the parameters of
supergravity GUTs.

\noindent
ACKNOWLEDGEMENTS

This research was supported in part by NSF Grant Nos. PHY-19306906 and
PHY-916593.

\vfill\eject
\vspace*{14.0cm}
\noindent
Fig. 1:  Maximum value of $\tau(p \rightarrow \bar\nu K^+)$ when
$m_{\tilde W_1} >$
100~GeV as a function of $m_0$, for the case $m_t =$ 150~GeV and $\mu >
0$ from
Ref.~\cite{ffi}.  The solid, dashed and dot-dashed lines correspond to
$m_{H_3}/M_G =
3,6$ and 10.  The lowest horizontal line is the current experimental
limit.  For
the two horizontal lines above, the lower and higher lines are the upper
bound for Super Kamiokande and ICARUS, i.e. the experiments are
sensitive to
lifetimes below these lines.
 \vfill\eject \vspace*{17.0cm}
\noindent
Fig. 2:  $BR(b \rightarrow s\gamma)$ in the SU(5) supergravity GUT for
$\mu > 0$
and $m_t =$ 150~GeV. \end{document}